\documentclass[12pt]{iopart}
\usepackage{color}
\usepackage{bm}
\usepackage{citesort}
\usepackage{graphicx}
\usepackage{amssymb}
\usepackage{tikz}

\def\be{\begin{equation}}
\def\ee{\end{equation}}
\def\bea{\begin{eqnarray}}
\def\eea{\end{eqnarray}}

\def\bi{\bibitem}

\def\ptl{\partial}

\usepackage{citesort}

\begin{document}
\nocite{*}

\title{Eicheons instead of Black holes}

\noindent \qquad June 20, 2020

\author{S.L. Cherkas\dag   \ and V.L. Kalashnikov\ddag
}

\address{\dag\
Institute for Nuclear Problems, Bobruiskaya 11, Minsk 220030,
Belarus}

\address{\ddag\ Facolt\'a
di Ingegneria dell'Informazione, Informatica e Statistica,
Sapienza Universit\'a di Roma, Via Eudossiana 18 00189 - Roma, RM,
Italia}

\begin{abstract}
A new spherically-symmetric solution for a gravitational field is
found in the conformally-unimodular metric. It is shown, that the
surface of the black hole horizon in the standard Schwarzschild
metric can be squeezed to a point by converting coordinates to the
conformally-unimodular metric. In this new metric, there is no
black hole horizon, while the naked singularity corresponds to a
point massive particle.  The reason for the study of this
particular gauge (i.e., conformally-unimodular metric)  is its
relation to the vacuum energy problem. That aims to relate it to
other physical phenomena (including black holes), and one could
argue that they should be considered in this particular metric.
That means the violation of the gauge invariance of the general
theory of relativity. As a result, the nonsingular ``eicheons''
\footnote{The term ``eicheon'' refers to the fundamental work
``Gravitation und Elektrizit$\ddot{\mbox{a}}$t'' by Hermann Weyl
where the concept of ``gauge field theory'' (``Eichfeldtheorie'')
was invented for the first time (e.g., see \cite{Trageser2018}).
We would emphasize by this term the decisive role of the gauge
conditions in our theory predicting an existence of extremely
compact but nonsingular astrophysical objects. Moreover, the
connotation with ``Eichel'' (that means an acorn in German)
implies that ``eicheon'' can have an internal structure and a
solid-like ``surface''.}
 appear as the non-point compact objects with
different masses and structures. They are a final product of the
stellar collapse, with the masses exceeding the
Tolman-Oppenheimer-Volkoff limit.
\end{abstract}

\section{Introduction}

One of the most intriguing objects in the theory of general
relativity (GR) is the ``black hole'' (BH) \cite{novikov,chandra},
which is a result of the collapse of astrophysical objects with
the masses exceeding the Tolman-Volkov-Oppenheimer (TVO) limit
\cite{tol,op}.  The gravitational waves registered recently are
considered as a result of the collision of massive BHs
\cite{waves}. Direct astrophysical observations also indicate the
extremely compact supermassive objects in the galactic cores
\cite{image} identified with BH. However, such BH evidences
should be considered with caution because they suggest only a
presence of some compact massive astrophysical object possessing
the BH properties for an external observer, but with the wholly
unknown internal structure.

The strange properties of BH forced many researchers (including A.
Einstein \cite {einstein}) to question the BH reality and consider
these objects as a pathological artifact of GR.   Several
discouraging facts are well-known:

1) The first issue is the presence of BH singularity with an
infinitely large density, which is physically questionable. In
order to avoid a singular state, the different modifications of GR
have been offered by taking into account torsion (see, for
example, \cite {pop}); space-time curvature limitations \cite
{muchanov}; or considering the gravitation as a physical tensor
field which requires gauge invariance violation and non-zero
graviton mass \cite{logunov}; and, at least, development of
quantum theories gravity, e.g., loop quantum gravity \cite {lqg}.
On the other hand, the BH singularity could be justified because
it is ``dressed,'' i.e., surrounded by a horizon, making it
invisible for an outside observer (the so-called ``cosmic
censorship'' principle \cite{censor}). The BH singularity found a treatment within 
the framework of GR (the concept of a so-called ``regular BH'' \cite{rgbh1,rgbh5}) through 
the modification of the energy conditions on the stress-energy tensor of matter \cite{rgbh3,rgbh2}. 
In particular, the impact of nonlinear electrodynamics and non-Abelian gauge fields on the BH formation and 
structure were studied \cite{rgbh4,rgbh6,rgbh8}, and the limits of the BH ``non-hair'' theorem were discussed \cite{rgbh7}.

2) The physical status of the ``event horizon'' itself could also
raise the questions. However, from GR, it is merely a ``one-sided
membrane'' (``no-return horizon'')  for the free-falling observer.
Nevertheless, a fact of the event horizon existence is doubted
both from classical and quantum viewpoints. For example, the
horizon formation relates to the stability of ultra-compact states
of a substance \cite{op}. The existence of such exotic stable
phases (e.g., free-quark phase \cite{quark}) could explain the
phenomenon of ultra-compact objects but with the size larger than
the horizon. Then, the concept of the event horizon, as well as
the unlimited gravitational collapse, are declared physically
meaningless in the field formulation of gravity with a massive
graviton \cite{logunov2,logunov3,logunov}. However, the
existence of ultra-compact objects, which are finely larger than
BH, is not disclaimed \cite{Vyblyi,kalash}. A quantum view on the
horizon issue reveals an ``information paradox,'' and a
"non-cloning" of quantum states, as well as the thermodynamical
problems \cite{th,hp,f,suss}.

Assuming the modernization of GR in its relation with the
``no-BH''-hypothesis refers to the synthesis of gravity with
quantum mechanics. That raises the question: what is the direction
of such modernization? In this regard, one can recall the known
statement by D.I. Blochintsev: ``Number of facts is always enough,
but fantasy is insufficient.'' \cite {bloch}.

The key fact indicating a possible path in the forest of the
alternative gravity theories is the vacuum energy problem. In GR,
any spatially uniform energy density (including that of zero-point
fluctuations of the quantum fields) causes the expansion of the
universe. Using the Planck level of UV-cutoff results in the
Planckian vacuum energy density $\rho_{vac}\sim M_p^4$ \cite{w},
which leads to the universe expanding with the Planckian rate
\cite {d1}. In this sense, the vacuum energy problem is an
observational fact \cite{arx}.

One of the possible solutions is to build a theory of gravity,
allowing an arbitrarily reference level of energy density. One
such theory has long been known. That is the unimodular gravity
\cite{a17,a21,a22,a23,unim}, which admits an arbitrary
cosmological constant. However, under using of the comoving
momentums cutoff, the vacuum energy density scales with time as
radiation \cite{arx}, but not as the cosmological constant.

Recently, another theory has been suggested \cite{ch1}, which
considers the Friedmann equation defined up to some arbitrary
constant. This constant corresponds to the invisible radiation
and, thus, can compensate the vacuum energy. In this case, one
could ask why the $\bm k$-cutoff of comoving momentums is used
instead of, for instance, a cutoff of physical momentums related
to $\bm p=\bm k/a$ ($a$ is the universe scale factor)? The answer
could be that it is relatively simple to construct a theory with
the $\bm k$-cutoff, but it is challenging to introduce the $\bm
p$-cutoff fundamentally. For instance, merely considering gravity
on a lattice gives rather fundamental theory with comoving
momentums restricted by the period of a lattice.

The next noteworthy fact of GR is the absence of a vacuum state,
which is invariant relative to the general transformation of
coordinates. It indicates the violation of gauge invariance at a
quantum level\footnote {As was found, most of the symmetries in
nature are violated. The exception is the color symmetry of the
quantum chromodynamics.}, but one could assume that the gauge
invariance should be broken at the classical level in GR, as well.
In particular, the five-vector theory of gravity (FVT) \cite {ch1}
assumes the gauge invariance violation in GR  by constraining the
class of all possible metrics in varying the standard
Einstein-Hilbert action. A question arises, how the classical
Schwarzschild solution looks in this class of metrics? The purpose
of this work is to elucidate the nature of compact astrophysical
objects in this limited class of conformally-unimodular metrics.

\section{Violation of gauge invariance in a framework of FVT}

The observational fact, that the bulk of vacuum energy density
does not affect the expansion of the universe, points out a
gravity theory, in which the reference level of energy density
could be chosen arbitrarily. Such a theory arises if one varies
the standard Einstein-Hilbert action over not all possible
space-time metrics $g_{\mu\nu}$, but over some class of
conformally-unimodular metrics\footnote{In this gauge, a
space-time metric is presented as a product of a common multiplier
by a 4-dimensional matrix with a determinant equal to -1,
including a 3-dimensional spatial block with unit determinant.}
\cite{ch1}
\be
\fl ds^2\equiv g_{\mu\nu} dx^\mu dx^\nu = a^2\left(1-\ptl_m
P^m\right)^2d\eta^2-\gamma_{ij} (dx^i+ N^i d\eta) (dx^j+
N^jd\eta),
\label{interv1}
\ee
where $x^\mu=\{\eta,\bm x\}$, $\eta$ is conformal time,
$\gamma_{ij}$ is a spatial metric, $a =\gamma^{1/6}$ is a locally
defined scale factor, and $\gamma=\det\gamma_{ij}$. The spatial
part of the interval (\ref{interv1}) reads as
\be
dl^2\equiv\gamma_{ij}dx^idx^j=a^2(\eta,\bm x)\tilde
\gamma_{ij}dx^idx^j,
\ee
where  $\tilde\gamma_{ij}=\gamma_{ij}/a^2$ is a matrix with the
unit determinant.

The interval (\ref {interv1}) is similar formally to the ADM one
\cite {adm}, but with the lapse function $N$ changed by the
expression $1-\ptl _ m P ^ m $, where $P ^ m $ is a
three-dimensional (relatively rotations) vector, and $\ptl _ m $
is a conventional particular derivative.

The starting point is the standard Einstein-Hilbert action
\cite{lan}
\bea
S_{grav}=-\frac{M_p^2}{12}\int \mathcal G\sqrt{-g}\,d^4x,
\label{sc}
\eea
where $\mathcal G=g^{\alpha \beta } \left(\Gamma _{\alpha \nu
}^{\rho }
   \Gamma _{\beta \rho }^{\nu }-\Gamma _{\alpha \beta
   }^{\nu } \Gamma _{\nu \rho }^{\rho }\right)$, and $M_p=\sqrt{\frac{3}{4\pi G }}=1.065\times 10^{-8}~kg$
   is the reduced Planck mass. The variation of (\ref{sc}) over vectors $\bm P$, $\bm N$ and  3-metric\footnote{Three dimensional spatial metric tensor can be written as the three-vectors triad. Thus 5-vectors appear in theory.}
$\gamma_{ij}$ leads to the FVT equations:
\bea
\frac{\ptl g^{\mu\nu}}{\ptl\gamma_{ij}}\left(\frac{(\ptl \mathcal
G\sqrt{-g})}{\ptl g^{\mu\nu}}-\ptl_\lambda\frac{\ptl (\mathcal
G\sqrt{-g})}{\ptl ( \ptl_\lambda g^{\mu\nu})}
-\frac{6}{M_p^2}T_{\mu\nu}\sqrt{-g}\right)=0,\nonumber\\
\frac{\ptl g^{\mu\nu}}{\ptl N^{i}}\left(\frac{\ptl (\mathcal
G\sqrt{-g})}{\ptl g^{\mu\nu}}-\ptl_\lambda\frac{\ptl (\mathcal
G\sqrt{-g})}{\ptl ( \ptl_\lambda g^{\mu\nu})}
-\frac{6}{M_p^2}T_{\mu\nu}\sqrt{-g}\right)=0,\nonumber\\
\frac{\ptl g^{\mu\nu}}{\ptl( \ptl_j P^{i})}\frac{\ptl}{\ptl
x^j}\biggl(\frac{\ptl (\mathcal G\sqrt{-g})}{\ptl g^{\mu\nu}}-
\ptl_\lambda\frac{\ptl( \mathcal G\sqrt{-g})}{\ptl ( \ptl_\lambda
g^{\mu\nu})}
-\frac{6}{M_p^2}T_{\mu\nu}\sqrt{-g}\biggr)=0,\label{last1}
\eea
where the energy momentum tensor  $T_{\mu\nu}=\frac{\delta
S_m}{\delta_{g^{\mu\nu}}}$ is introduced. The last equation
(\ref{last1}) is weaker than the corresponding Hamiltonian
constraint of GR. On the other hand, the restrictions $\bm
\nabla(\bm \nabla\cdot \bm P)=0$, $\bm \nabla (\bm \nabla\cdot \bm
N)=0$ on the Lagrange multipliers arise in FVT. Using the gauge $\bm \nabla \cdot \bm N=0$ provides the
Hamiltonian constraint fulfillment up to some constant \cite
{ch1}.

\section {A spherically symmetric static gravitational field}

The spherically symmetric metrics belonging to the class
(\ref{interv1}) reads as:
\begin{equation}
\fl ds^2 = a^2(d\eta ^2 - \tilde {\gamma }_{ij} dx^idx^j) = e^{2\alpha
}\left( {d\eta ^2 - e^{ - 2\lambda }(d{\bm x})^2 - (e^{4\lambda }
- e^{ - 2\lambda })({\bm x}d{\bm x})^2 / r^2} \right),
\label{eq31}
\ee
where  $r = \vert \bm x \vert $, $a = \exp \alpha $, $\lambda $
are the functions of $\eta , r$.  The matrix  $\tilde \gamma_{ij}$
with the unit determinant is expressed through $\lambda(\eta,r)$.
Thus, for the spherically symmetric case, the equations
(\ref{last1}) take the form
\bea
 \fl{\mathcal H} = e^{2\alpha }\biggl(  - \frac{1}{2}\alpha ^{'2} +
\frac{1}{2}\lambda ^{'2} - \frac{e^{2\lambda }}{6r^2}
+\frac{e^{2\alpha}}{{M_p^2}}\rho+ e^{- 4\lambda }\biggl(
\frac{1}{6r^2} - \frac{4}{3}\partial _r \alpha {\kern 1pt}
\partial _r \lambda + \frac{1}{6}\partial _r \alpha ^2 +\nonumber\\
\frac{2\partial _r \alpha }{3r} + \frac{1}{3}\partial _{r,r}
\alpha + \frac{7}{6}\partial _r \lambda^2 - \frac{5\partial _r
\lambda }{3r} - \frac{1}{3}\partial _{r,r} \lambda  \biggr)
   \biggr) = const,
\label{eq32}
\eea
\begin{equation}
\fl \mathcal P= e^{2 \alpha}\bigl( -\partial_r \alpha \left(
{\alpha^\prime + 2\lambda^\prime} \right) - \partial_r
\lambda^\prime+ \partial_r \alpha^\prime - \left( {3/r -
3\partial_r \lambda } \right)\lambda^\prime\bigr) = 0,
\label{eq33}
\end{equation}
\bea
 \fl \alpha ^{\prime\prime} + \alpha ^{\prime2} + \lambda ^{\prime 2} = e^{ - 4\lambda }\biggl(  -
4\partial _r \alpha {\kern 1pt} \partial _r \lambda + \partial _r
\alpha ^2 + \frac{2\partial _r \alpha }{r} + \partial _{r,r}
\alpha +
 \frac{7}{3}\partial _r \lambda ^2 -\nonumber \\ \frac{10{\kern 1pt}
{\kern 1pt}
\partial _r \lambda }{3r} - \frac{2}{3}\partial _{r,r} \lambda +
\frac{1}{3r^2}\left( {1 - e^{6\lambda }} \right)
\biggr)+\frac{e^{2\alpha}}{M_p^2}(3p-\rho),
\label{eq34}
 \eea
\bea
\fl \lambda ^{\prime\prime} + 2\alpha ^\prime\lambda ^\prime =
\frac{2}{3}e^{ - 4\lambda }\biggl(  -
\partial _r \alpha {\kern 1pt} \partial _r \lambda - \partial _r \alpha ^2 +
\partial _{r,r} \alpha + \partial _r \lambda ^2 - \frac{1}{2}\partial _{r,r}
\lambda - \frac{1}{r}\partial _r \alpha\nonumber\\ -
\frac{1}{r}\partial _r \lambda + \frac{1}{2r^2}\left( {e^{6\lambda
} - 1} \right) \biggr),
\label{eq35}
\eea

\noindent where prime denotes differentiation over  $\eta $. Eq.
(\ref {eq32}) is the Hamiltonian constraint, but it includes an
arbitrary constant now. If this constant equals zero, one returns
to GR. Eq. (\ref {eq33}) follows from the momentum constraint. The
expressions (\ref {eq34}), (\ref {eq35}) are the equations of
motion.

Differentiation of the constraints over time $\eta $ results in
the following equations
\bea
\mathcal H^\prime=\frac{1}{3r^2}\ptl_r\left(e^{-4\lambda}r^2\mathcal P\right),\label{s1} \\
 \mathcal P^\prime=\ptl_r \mathcal H,\label{s2}
\eea
which are satisfied if the equations of motion (\ref{eq34}),
(\ref{eq35}) are fulfilled, and, besides, the following equations
for the energy density and pressure are enforced:
\be
\rho^\prime+3(p+\rho)\alpha^\prime=0,~~~~\ptl_r
p+(p+\rho)\ptl_r\alpha=0. \label{mat}
\ee

In GR, the equations (\ref {mat}) arise from the Bianchi
identities resulting in $D ^\mu T_{\mu\nu}=0$. In the FVT case,
the relations (\ref {mat}) arise from the requirement of the
constraints conservation in the time (\ref{s1}), (\ref{s2}).
Generally, the equations (\ref{s1}), (\ref {s2}) satisfy not only
$\mathcal H = 0 $, $\mathcal P = 0 $, as in GR, but weaker
conditions $\mathcal H = const $, $\mathcal P = 0 $, as reflected in Eq. (\ref{eq32}).  A constant on the right hand side of
Eq. (\ref {eq32}) compensates the bulk of the vacuum energy, and,
after the compensation (if it is exact), the equations become the
same as in GR. All this take a place in the conformo-unimodular
metric (\ref{eq31}), in which we will find the  Schwarzschild
solution, assuming  the time derivatives, as well as pressure and
density equal to zero in Eqs. (\ref {eq32}--\ref{eq35}).
Expressing the derivatives $\partial _{r,r} \lambda $, $\partial
_{r,r} \alpha $ from Eqs. (\ref{eq34}), (\ref{eq35}) and
substituting them into  (\ref{eq32}) under the $const = 0$, one
finds
\be
 - 3r^2\left( {\frac{d\alpha }{dr}} \right)^2 + 4r\frac{d\alpha }{dr}\left(
{r\frac{d\lambda }{dr} - 1} \right) - \left( {r\frac{d\lambda
}{dr} - 1} \right)^2 + e^{6\lambda } = 0.
\label{eq37}
\ee
To obtain a solution of the equations
 (\ref{eq34}), (\ref{eq35}), (\ref{eq37}), let us make the following
 substitution
\begin{equation}
\lambda = \alpha + \ln \left( {\left( {1 - e^{2\alpha }} \right)r
/ r_g } \right),
\label{eq38}
\end{equation}
where the Schwarzschild radius is introduced for the sake of
dimensionless of the expressions under logarithm.
\begin{figure}[htbp]
\centerline{\includegraphics[width=8cm]{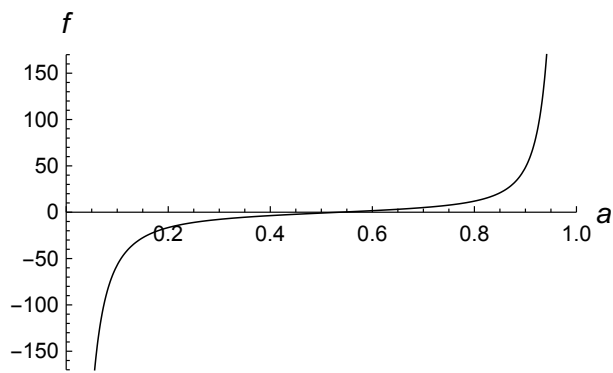}} \caption{Plot of
the function   $f(a)$ defined by Eq. (\ref{f}). }
\label{fig22}
\end{figure}
 As a result, Eq. (\ref{eq37}) takes the form:
\begin{equation}
r^4e^{4\alpha }\left( {e^{2\alpha } - 1} \right)^8 - 4\left(
{\frac{d\alpha }{dr}} \right)^2r_g^6 = 0
\label{eq39}
\end{equation}
and has the solution
\begin{equation}
\label{eq40}
\alpha (r) = \ln \left( {f^{ (- 1)}\left( {\frac{r^3 - r_0^3
}{6r_g^3 }} \right)} \right),
\end{equation}
where $f^{ - 1}$ is the inverse function of
\be
 f(a) = 2\ln \left( {\frac{a^2}{1 - a^2}} \right) + \frac{30a^4
- 12a^6 - 22a^2 + 3}{6a^2\left( {a^2 - 1} \right)^3} \label{f}
\ee
and  $r_0$ is an integration constant. The function $f(a)$, which
maps an interval $(0,1)$ into $\mathbb{R}$,  is mutually
single-valued function  shown in Fig. \ref{fig22}. Using
(\ref{eq40}), (\ref{f}) and the rules of the differentiation of
the inverse function allows calculating
\be
\fl \frac{d\alpha}{dr}=\frac{r^2}{2 r_g^3 f^{(-1)}\left(y\right)
f'\left(f^{(-1)}\left(y\right)\right)}=\frac{r^2 f^{(-1)}(y)^2
\left(f^{(-1)}(y)^2-1\right)^4}{2 r_g^3}=\frac{8 e^{6 \alpha } r^2
\sinh ^4\alpha }{r_g^3},
\label{der1}
\ee
where $y=\frac{r^3 - r_0^3 }{6r_g^3 }$. Similar calculations give
\bea
\frac{d\lambda}{dr}=\frac{8 e^{6 \alpha  } r^2 \sinh ^4\alpha
(\coth \alpha
 +2)}{r_g^3}+\frac{1}{r},\label{der2}\\
\frac{d^2\alpha}{dr^2}=\frac{16}{r_g^6} e^{6 \alpha } r \sinh
^4\alpha  \left(8 e^{6 \alpha } r^3 \sinh ^3\alpha  (3 \sinh
\alpha +2 \cosh \alpha )+r_g^3\right),\label{der3}\\
 \frac{d^2\lambda}{dr^2}=
 \frac{64 e^{12 \alpha } r^4 \sinh ^6\alpha  (7 \sinh (2 \alpha )+8 \cosh (2 \alpha )-5)}{r_g^6}-
  \frac{1}{r^2}+\nonumber\\\frac{16 e^{6 \alpha } r \sinh ^4\alpha  (\coth \alpha
  +2)}{r_g^3}.\label{der4}
\eea
Substitution of Eqs. (\ref{der1}), (\ref{der2}), (\ref{der3}),
(\ref{der4}) into Eqs. (\ref{eq32}), (\ref{eq33}), (\ref{eq34}),
(\ref{eq35}) demonstrates that the lasts are satisfied at $p =\rho
= 0 $, and $const = 0 $. Thus, Eqs. (\ref {eq38}), (\ref {eq40}),
(\ref {f}) are the exact spherically-symmetric static solution of
the Einstein equations in vacuum. From the physical viewpoint, it
appears that  $const$ in Eq. (\ref {eq31}) compensates a vacuum
energy of quantum fields.

The function $\alpha$ is not singular everywhere, as it is shown
 in Fig. \ref{fig1} (a), whereas the   function
$\lambda$, describing the deviation of conformally-unimodular
metric geometry from the Schwarzschild one, is singular only at
$r=0$.  
\begin{figure}[htbp]
\centerline{\includegraphics[width=15cm]{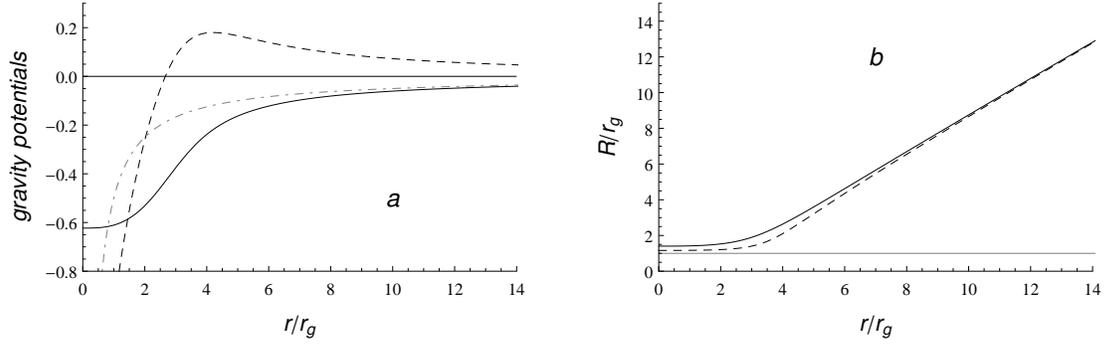}} \caption{(a)
"Gravitational potentials"$\,$ describing the metric (\ref
{eq31}). Solid and dashed lines correspond to $\alpha (r)$, and
$\lambda (r)$, respectively. The dashed-doted line is the
Newtonian potential $\varphi=-\frac {r_g} {2r}$. (b) Coordinate
transformation $R(r)$ mapping the metric (\ref {eq31}), (\ref
{eq42}) to the canonical Schwarzschild form (\ref {eq41}) for the
different integration constant values $r _0$ in the expression
(\ref{eq40}). Solid and dashed lines correspond to  $r_0=0$ and
$r_0=3r_g$, respectively. The level of $R=r_g$ is marked by the
gray horizontal line.}
\label{fig1}
\end{figure}

Let us compare the solution (\ref{eq38}),  (\ref{eq40}) with the
canonical Schwarzschild one which is \cite{lan}
\begin{equation}
\label{eq41}
ds^2 = (1 - r_g/R)dt^2 - (1 - r_g / R)^{-1}dR^2 -
R^2(d\theta^2+\sin^2\theta d\phi^2 ).
\end{equation}

For this aim, we rewrite the interval
 (\ref{eq31}) in the spherical coordinates
\begin{equation}
ds^2 = e^{2\alpha }\left( {d\eta ^2 - dr^2e^{4\lambda } - e^{ -
2\lambda }r^2\left( {d\theta ^2 + \sin^2\theta d\phi ^2} \right)}
\right){\kern 1pt} {\kern 1pt}.
\label{eq42}
\end{equation}
\begin{figure}[htbp]
\centerline{\includegraphics[width=12.51cm]{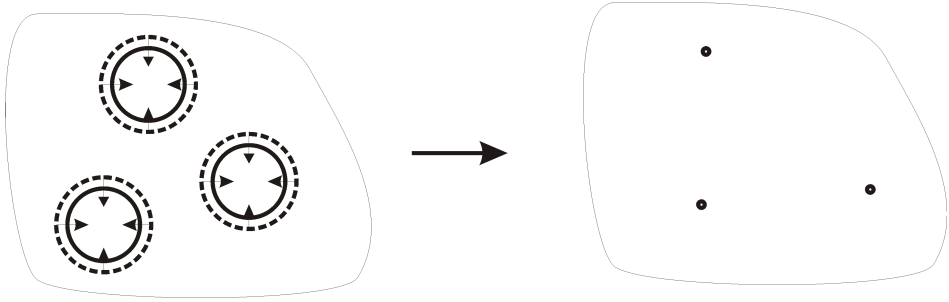}}
\caption{``Squeezing'' of the BHs of Schwarzschild horizons into
the nodes with point masses.}
\label{fig2}
\end{figure}
The solutions (\ref {eq41}) and (\ref {eq42}) of the Einstein
equations should be interrelated by the transformation of
coordinates $t =\eta $, $R = R (r) $, which gives another way to
deduce Eqs. (\ref{eq38}), (\ref{eq39}). Actually, equating the
coefficients at $dt^2=d\eta^2$, $d\theta ^2$ and $\sin^2\theta d\phi
^2$, as well as the radial terms in the intervals (\ref {eq41}),
(\ref {eq42}) gives the equations
\bea
1 - r_g/R=e^{2\alpha },~~~~~~~~~~~~~~~~~~~\label{e1}\\
R^2=r^2e^{-2\lambda+2\alpha},~~~~~~~~~~~~~~~~\label{e2}\\
 (1 -
r_g/R)^{-1}\left(\frac{dR}{dr}\right)^2=e^{4\lambda+2\alpha}\label{e3}.
\eea
The relations (\ref {e1}), (\ref {e2}) result in the expressions
(\ref{eq38}) and $R(r) = r_g\left(1-e^{2\alpha}\right)^{-1}$,
which give (\ref {eq39}) after substitution in (\ref {e3}). As is
shown in Fig. \ref {fig1} (b), the solutions (\ref {eq38}), (\ref
{eq40}) describe only a part of the external space $r_g^{+}$ of
the Schwarzschild solution by virtue of $\lim_{r\to 0} R
(r)\gtrsim r_g$. Fig. \ref {fig2} illustrates this fact in the
following way. Let there is a space filled with the Schwarzschild
BHs. Then, by inverse coordinates transformation having the form
$r (R)$ in the vicinity of each hole, one can squeeze holes into
the nodes $r=0$ and consider that point particle is placed in each
node. The space-time obtained in such a way will represent a
single causally connected region. That corresponds to the existence of impenetrable surface at $R(0)>r_g$ in 
the metric  (\ref{eq41}) . This situation resembles that in theory with massive graviton, where the physical singularity corresponds to $R=r_g$ \cite{logunov3,kalash}. 

In contrast to a regular BH concept admitting a horizon
without a singularity \cite{rgbh3,rgbh2,rgbh4,rgbh6,rgbh8}, the conformally unimodular gauge
changes the terminology completely. Namely, all its solutions have no
horizon without requiring any exotic states of matter.

In principle, using the Dirac delta function and writing the
density energy in Eq. (\ref {eq32}) as $\rho (\bm
x)=e^{-3\alpha}\delta^{(3)}(\bm x)$ \cite {ch1}, one could
consider the solutions (\ref {eq38}), (\ref {eq39}) as
corresponding to the $\delta$-source, but such a consideration is
rather formal because the equations of gravity are nonlinear,
whereas  the product of generalized functions cannot be defined
correctly \cite{vlad}. Some additional definition of the structure
of the Dirac delta function is required to overcome this
difficulty \cite {kat}. For instance, one could consider a
physical model of delta-function in the form of a sphere of
constant density, with the radius approaching zero along with the
density tending to infinity.

\section {Compact objects of the constant density}

\subsection {Uniform compact object in the Schwarzschild metric}

The well-known Tolmen-Oppenheimer-Volkov equation (TOV) \cite{op},
which defines the maximal mass of a stable neutron star, written
in the  Schwarzschild type metric
\be
ds^2=B(R)dt^2-A(R)dR^2-R^2d\Omega,
\label{ab}
\ee
reads as:
\begin{equation}
\fl p^\prime(R)=-\frac{3}{4\pi M_p^2R^2} \mathcal M(R) \rho (R)
\left(\frac{1+4 \pi R^3 p(R)}{\mathcal M(R)}\right)
\left(1+\frac{p(R)}{\rho (R)}\right)\left(1-\frac{3 \mathcal
M(R)}{2\pi M_p^2 R}\right)^{-1},
\label{op}
\end{equation}
where the function $\mathcal M(R)=4\pi \int_0^R \rho(R^\prime)
R^{\prime 2}dR^\prime$.

Although an ideal incompressible fluid seemed to be not existing
in nature, an approximation of constant density \cite{wein} allows
describing the general features of the compact physical objects.
In this case  $\mathcal M(R)=\frac{4\pi}{3}\rho R^3$, the solution
of Eq. (\ref{op}) takes the form

\be
 p(R)=\rho\,\frac{  \sqrt{M_p^2-2 \rho
\, R^2}-\sqrt{M_p^2-2 \rho\, R_f^2}}{3\sqrt{M_p^2-2 \rho\, R_f^2}-
\sqrt{M_p^2-2\rho  R^2}},
\label{rop}
\ee
where $R_f $ is the radius of an object. As it is seen from the
formula (\ref {rop}), pressure turns to infinity at $R =\sqrt {4M
_p ^2/\rho-9R_f^2} $, that points to some limitations on the size
of the object.  A condition of pressure finiteness yields
$4M_p^2/\rho<9R_f^2$, i.e., the size of an object has to be
$R_f>\frac{9}{8}r_g$, where $r_g=\frac{3m}{2\pi M_p^2}$,
$m=\mathcal M(R_f)=\frac{4\pi\rho R_f^3}{3}$.

\subsection {Shell compact object in the Schwarzschild metric}

Let us consider a more complex model of astrophysical object
consisting of two immiscible and incompressible liquids with the
densities  $\rho_1 $ and $\rho_2$. It is the simplest prototype
for the neutron star with a non-uniform internal structure \cite
{ns,ns2}.

Then, the function $\mathcal M(R)$ is written as
\be
 \mathcal M(R)= \frac{4\pi}{3} \left\{ {{\begin{array}{*{20}c}
\rho_1R^3,~~~~R<R_i,\\
\rho_2(R^3-R^3_i)+\rho_1R_i^3,~~~~~R_i<R<R_f,\\
 \rho_2(R_f^3-R^3_i)+\rho_1R_i^3,~~~R>R_f.
\end{array}} } \right.
\label{m2}
\ee

When $\rho_1$ is close to zero and $\rho_2=\rho$, the function
(\ref{m2}) becomes
\be
 \mathcal M(R)= \frac{4\pi \rho}{3} \left\{ {{\begin{array}{*{20}c}
0,~~~~R<R_i,\\
R^3-R^3_i,~~~~~R_i<R<R_f,\\
 R_f^3-R^3_i,~~~R>R_f.
\end{array}} } \right.
\ee
The analytical solution of (\ref{op}) for pressure with this
$\mathcal M(R)$ is still cumbersome, however calculation shows
softer condition for the pressure finiteness, which is shown in
Fig. \ref {shell}. For a sufficiently thin shell  $R_f$ approaches
to $r_g$.

\begin{figure}[htbp]
\centerline{\includegraphics[width=8.51cm]{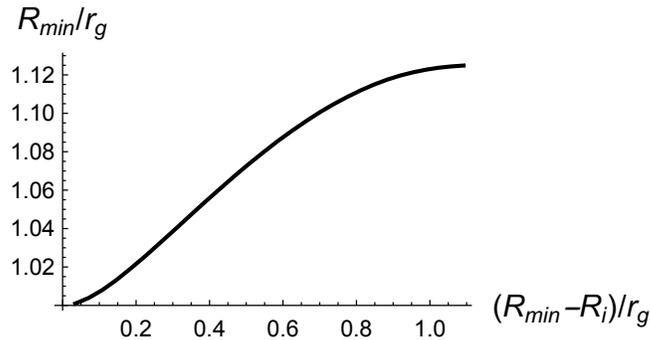}} \caption
{Minimum possible outer radius $R_f=R _{min}$ in dependence on the
thickness of a  shell  in the Schwarzschild metric (\ref{ab}).}
\label{shell}
\end{figure}

\subsection {Compact object in the conformally-unimodular metric}

\subsubsection {Object of a star class}

Modern observations of ultra-compact BH-like objects, formed as a
result of collapse of massive stars, give the maximum estimation
of their masses of order of $m = 15\div 36\, m_{\bigodot}$ \cite
{bh1,bh2}. Let us consider the constant density objects in the
metric (\ref {ab}) related by the coordinate transformation  $R
(r)=\exp\left (\alpha(r) - \lambda(r)\right) $ with the metrics
(\ref{eq38}). A quantity  $r_f$ denoting boundary of a matter in the conformally-unimodular metric
corresponds to $R_f=R(r_f)$ in the
metric of (\ref {eq42}), while $R_i=R(0)$. Because the horizon is
absent in this metric, nothing prevents $r_f$ to be smaller then
$r_g$. The functions $\alpha(r)$, $\lambda(r)$, $p(r)$ are defined
by the equations (\ref{eq32})-(\ref{eq35}) within a sphere
occupied by matter. The boundary conditions at $r=r_f$ are given by
linkage with the Schwarzschild solution (\ref {eq38}),
(\ref{eq40}).

After solving of Eqs. (\ref {eq32}) - (\ref {eq35}), the mass of
an object can be recovered
\be
m=\frac{4\pi}{3}\rho\left(R_f^3-R_i^3\right)=4\pi\rho\int_{0}^{r_f}
e^{3 (\alpha (r)-\lambda (r))} \left(r \frac{d\alpha}{dr}-r
\frac{d\lambda}{dr} +1\right)r^2dr, \label{mass}
\ee
which determines the Schwarzschild radius
$r_g=\frac{3}{2\pi}\frac{m}{M_p^2}$, appearing in the formulas
$(\ref{eq38})$ and $(\ref{eq40})$.
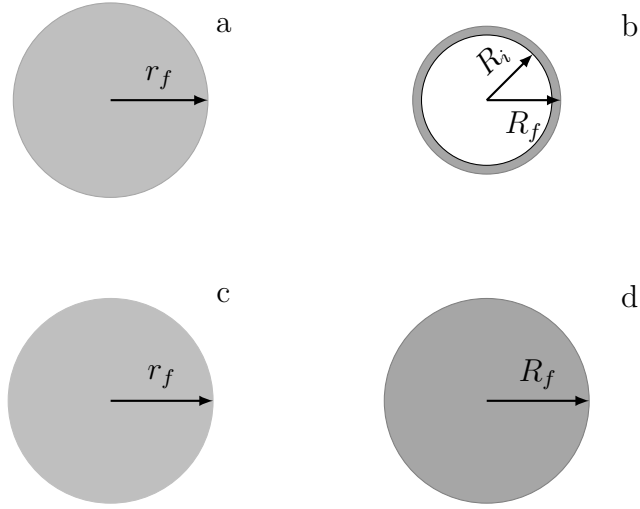
\begin{figure}[th]
\centering
\begin{tikzpicture}
\draw[gray!70,fill=gray!50] (0,4) circle (1.29379cm);
\draw[gray,fill=gray!70] (5,4) circle (0.98358cm);
\draw[black,fill=white] (5,4) circle (0.867064cm);
\draw[gray,fill=gray!70] (5,0) circle (1.36188cm);
\draw[gray!50,fill=gray!50] (0,0) circle (1.36188cm);
  \draw (1.5,5) node {a};
  \draw (6.9,5) node {b};
  \draw (1.5,1.4) node {c};
  \draw (6.9,1.4) node {d};
\draw[->,  thick,  arrows={-latex}]  (0,0) -- (1.36188,0)
node[sloped,midway,above=0.4] {$r_f$};
 \draw[->,  thick,   arrows={-latex}]  (5,0) -- (6.36188,0) node[sloped,midway,above] {$R_f$};
 \draw[->,  thick,  arrows={-latex}]  (0,4) -- (1.29379,4) node[sloped,midway,above] {$r_f$};
 \draw[->,  thick,  arrows={-latex}]  (5,4) -- (5.98358,4) node[sloped,midway,below] {$R_f$};
 \draw[->,  thick,  arrows={-latex}]  (5,4) -- (5.61311,4.61311) node[sloped,midway,above] {$R_i$};
\end{tikzpicture}
\caption {\normalfont (a) A  compact object of uncompressible
fluid ($\rho_0=0.43\,M_p^2r_g^{-2}$) with the radius $r_f=2r_g$ in
the conformally-unimodular metric (\ref {eq42}) looks as a shell
(b) with the boundaries  $R _i=R(0)=1.34\,r_g$ and $R_f=R (r_f)
=1.52\,r_g$ in the Schwarzschild type metric (\ref {ab}).\\
(c),(d) Low density object $\rho_0=5.0117\times 10^{-10}\,
M_p^2r_g^{-2}$ looks as a solid ball $R_f\approx r_f=1000 r_g$ in
both metrics if parameter $r_0$ in (\ref {eq40}) equals
$r_0=-96.75$.}
\label{fig31}
\end{figure}

\begin{figure}[htbp]
\centerline{\includegraphics[width=8.51cm]{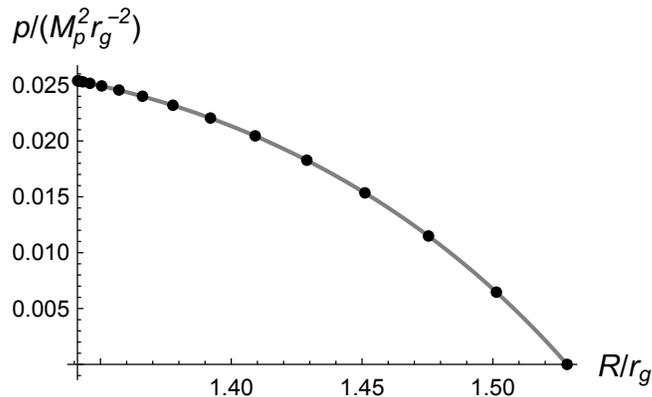}} \caption {The
pressure obtained by solving the TOV equation (points) and the
equations (\ref {eq32}), (\ref {eq33}), (\ref {eq34}),
(\ref{eq35}) (solid curve). The values of parameters correspond to
Fig. \ref{fig31} (a),(b). }
\label{fig32}
\end{figure}

Let us first discuss compact objects in the metric (\ref {eq31}),
(\ref {eq42}) where the matter occupies a sphere with the size
less or an order of the Schwarzschild radius (see Fig. \ref
{fig31}), (a), (b). As could be expected, the potential $\alpha $,
which was finite in the case of a point source, remains finite.
The potential $\lambda$, which was infinite in the point where the
point-like source was located, becomes finite inside a uniformly
mass distribution within a ball.

The internal structure of a compact object in the metric (\ref
{ab}) is defined by the internal $R_i$ and external  $R_f$ radii,
while there is only single external radius $r_f $ in the metric
(\ref {eq42}), and, besides, there is an additional parameter
$r_0$ in the external Schwarzschild solution (\ref{eq38}), (\ref
{eq40}). Thus, the meaning of this additional parameter $r_0 $
becomes clear. Namely, it defines the internal structure of an
object.  It is not surprising that partial information about the
pseudo-BH structure is contained in the Schwarzschild external
solution in the form of parameter $r_0$ because no real BH in the
conformally-unimodular metric exists.

Certainly, the pressure $P(R)$, $R\in\{R _i, R_f\}$ obtained by
the solution of the TOV equation matches the pressure recovered
from Eqs. (\ref {eq34}), (\ref {eq35}), (\ref {mat}) in the
parametric form $p(r),R (r)$, $r\in\{0,r_f\}$  as it is shown in
Fig. \ref {fig32}.

\subsubsection{Supermassive object }

Recently, the existence of supermassive compact objects in galaxy
nuclei was confirmed, and their masses were estimated as $m =
6.5\times 10^9\,m_{\bigodot}$ \cite{image}. Assuming the existence
of some maximal density in nature  $\rho _ {max}\simeq M_p^4$,
after conversion to the units $M_p ^ {2} r_g^{-2}$, results in
$\rho_0=\rho_{max}=3.4\times 10^{95}M_p^{2} r_g^{-2}$. For the
conformally-unimodular metric, the size of this object turns out
to be very small and, as calculations show, the potentials
$\lambda$ and $\alpha$ inside a ball can be estimated by taking
expressions (\ref {eq38}), (\ref {eq40}) for empty space (i.e.,
the boundary conditions affect $\alpha(r)$ stronger than the
``structure'' of an object). Moreover, one has at a small $r/r_g $
\be
\alpha(r)\approx\ln\left(a_0(r_0)+\frac{r^3-r_0^3}{6r_g^3
k(r_0)}\right),
\label{appr}
 \ee
because the value of $a$ tends to some constant $a_0$  at
$r\rightarrow 0 $. The parameters $a_0(r_0)$ and $k(r_0)$ are the
functions of $r_0$. The expression (\ref {appr}) has been derived
by the expansion of the function $f (a) $ into Taylor's series at
the point  $a_0$ up to the first order in $a-a_0$. After this
expansion finding of the inverse function $f^{(-1)}$ becomes
elementary. The value of $a_0$ is a root of the equation $f
(a)-\frac {r _0^3} {6r _g^3} = 0$ and $k=f^\prime(a_0)$. Further,
as an example, $r_0=0$ will be considered, when $a_0\approx0.54$,
and $k=25.2$.

The calculation of mass using (\ref {mass}) yields
\be
m\approx \frac{4 \pi}{3}\rho_0\frac{a_0}{k(1-a_0^2)^4}{r_f}^3,
\ee
 giving the estimation for
$r_f\approx 2.6\times 10^{-32} r_g=1.5 \times 10^{16}M_p^{-1}$.
 The radius of a boundary surface in
the Schwarzschild metric can be approximated from (\ref{e1}),
(\ref{appr})
\be
R(r)\approx\frac{r_g}{1-\left(a_0+\frac{r^3}{6 r_g^3 k}\right)^2},
\ee
which gives $R_i=R(0)\approx 1.4 r_g$.

As was already mentioned, it is possible to ``approach'' closer to
the Schwarzschild radius if to take another value of parameter $r
_ 0 $. The thickness of surface $\Delta R=R_f-R_i\approx
|\frac{dR(r)}{dr}|_{r\rightarrow r_f}\approx 7.5\times
10^{-97}r_g\approx 4.3 \times 10^{-49} M_p^{-1}$. So small
thickness $\Delta R $ of surface results from the hugeness of its
area. The second equation of (\ref{mat}), using (\ref{appr}) and
setting boundary condition $p( r_f)=0$ allow estimating the
pressure
\be
p(r)\approx\frac{r_f^3-r^3}{r^3+6a_0 k \,r_g^3}\,\rho_0.
\ee
The maximum of the pressure is $p\approx 0.07 M_p^{2} r_g^{-2}$,
i.e., it is much lower than the density $\rho_0 $, due to low
potential gradient $\alpha (r) $ inside an ``eicheon'' given by
Eq. (\ref {appr}), or from the extremely small surface thickness
$\Delta R$ in the terms of the TOV approach.
 This situation is analogous
that in theory with massive graviton, where impenetrable surface exists before
 $r_g$ \cite{logunov3,kalash} in the metric of (\ref{ab}) type. Since the
real astrophysical objects must have the radii $>r_g$,
this lifts the issue of the BH singularity in \cite{logunov3,kalash}.

\subsection{Low density objects}

Low density objects (recall, for example, that the sun radius is
$R _ f\approx 236000 r_g$) illustrated in Figs. \ref {fig31}, (c),
(d), which represents a solid ball in the Schwarzschild metric
(\ref {ab}).
 Although they are not related to the compact objects but
 could be considered for the completeness of the picture. It turns out
 to be that, in this case, the value of parameter $r_0$ in external metric (\ref {eq40})
 is fixed by the requirement $r=0$ when $R=0$. As is shown in Fig. \ref{fig6}, the condition of $R=0 $
 at $r=0$ meets only if $r_0=-96.75$ for a non-compact object of the radius of
$r_f=1000r_g$. Then the value of $R(r)$ becomes almost the same
with the $r$, as it is shown in Fig.  \ref{fig6}.  Thus,
one may conclude that the ``friable'' objects can also be
described consistently in the conformally-unimodular metric.
\begin{figure}[htbp]
\centerline{\includegraphics[width=8.51cm]{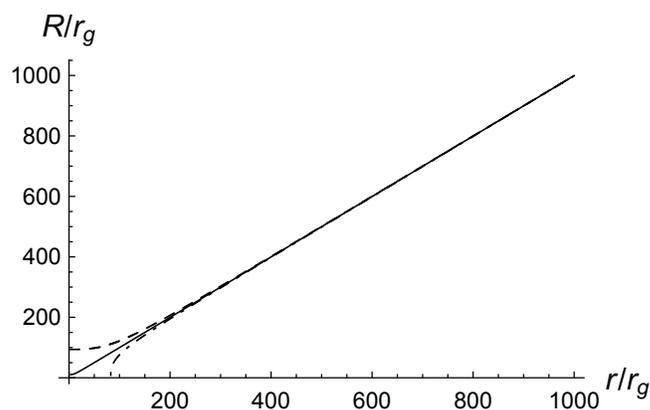}} \caption
{Dependence of  $R(r)$ for a ball of $R_f\approx r_f=1000\,r_g$
filled with the ``friable'' matter at different values of the
parameter $r_0$ in the external metric: $r_0 = -120 $ (dashed
line), $-96.75$ (solid), $-70$ (dash-dotted). Density of matter is
of $\rho_0\approx 5\times 10^{-10}\,M_p^2r_g^{-2}$. }
\label{fig6}
\end{figure}

In this context, it is interesting to imagine a low density
object, but with an empty core surrounded by a firm ``artificial
surface'' composed from an incompressible liquid, for instance.
Such an object can also be described in the conformally-unimodular
metric, with the $r$-coordinate running from 0 to $r_f $ and $r_0$
should have a small negative value, that corresponds to an
internal cavity in the Schwarzschild coordinates.

The difference between the ``friable'' and compact dense objects
is, that  for the first one, internal cavity in the Schwarzschild
metric could eliminate an  by taking a larger value of $r_0 $ in
the metric (\ref {eq42}). In contrast, for a dense object, the
cavity in the Schwarzschild metric cannot be eliminated in any
way.

\section{Objects made of dust}

Let's consider the motion of a sample dust particle in the metric
(\ref {eq42}) in the neighborhood of $r=0$, where
\be
\alpha \approx const, ~~~~~~ \lambda = \alpha + \ln \left( {\left(
{1 - e^{2\alpha }} \right)r / r_g } \right) \approx const + \ln
r.\ee

The radial geodesics satisfy the equation
\be
\ddot {\eta } = 0, ~~~~ \ddot {r} + 2\dot {r}^2 / r = 0,
\label{rad}
\ee
where a dot denotes a derivative over the proper time $ s$. The
solution of Eq. (\ref {rad})
\be r(\eta ) = r_{in}^{2 / 3} \left( {r_{in} -
3v{\kern 1pt} (\eta - \eta _{in} )} \right)^{1 / 3}
\ee
shows that the sample particle, placed initially at the point
$r_{in}$, $\eta =\eta_{in} $ and having the speed $v$ directed
towards center, reaches the point $r=0$ for the finite time.

Qualitatively, the formation of objects with the equation of state
of the dust type, i.e., having very low pressure, can be imagined
as the radial falling of the dust particles in the ``eicheon''
field. As a result, dust particles  are accumulated in the
vicinity of $r< r_g$, where the gradient of potential $\alpha$ is
negligible, i.e., the gravitational field is absent. In this
conformally-unimodular metric, ``eicheon'' is similar to a trap
because a particle needs to overcome the region of large potential
to escape from such a trap. This picture is quite similar to those
discussed in Ref. \cite{muchanov}.

On the other hand, in the Schwarzschild type metric (\ref{ab}), a layer, where the
dust particles are accumulated,  is very thin. The thickness is
determined by the residual pressure if to consider that some small
pressure is still present. This picture resembles a a very thin surface
discussed in \cite{logunov,Vyblyi,kalash,kalash2}, where it
originates from the non-zero mass of graviton.

\section{Conclusion}

We considered the conformally-unimodular gauge, which was chosen
for the sake of avoiding the problem of vacuum energy. A
requirement that the bulk vacuum energy $\rho_{vac}\sim M_p^4$
does not influence the curvature of space-time leads to the gauge
invariance violation and restricts the class of the possible the
metrics.
 That results in the absence of BH and the appearance of ``eiheons'' instead. All the
 compact real astrophysical objects in this class of the metrics look like solid balls of different sizes without any singular surfaces
(``horizons''). If such the compact objects $r_f\leq r_g$ are
considered in the Schwarzschild metric, they look like a matter
layer distributed over the impenetrable spherical shell with a
radius greater than the Schwarzschild one.

We have considered only spherically symmetric solutions in the framework of the FVT, and see a further generalization of the presented model by taking into account the ?Eicheon? spinning and its ``non-hair'' properties. Preliminary results concerning Kerr solution is sketched in the Appendix.

\appendix
\section{The Kerr's solution into the conformally-unimodular metric}

Whereas the static and spherically symmetric ?eicheon?? was above considered, the real astrophysical objects are spinning in nature. As it is well-known, the rotating BH has a region outside the Schwazshcild sphere known as the ergosphere \cite{chandra}, which plays a fundamental role in such phenomena as the Lense-Thirring (or frame-dragging) effect \cite{lense}, the particle acceleration around a rotating black hole, and the Penrose process (i.e., the energy extraction from a rotating black hole) \cite{bhat}.

Therefore it is interesting to express the Kerr solution in the
unimodular metric. The Kerr solution in the form of
Boyer-Lindquist is written as \cite{chandra,boyer}:
\bea
\fl ds^2=\left(1-\frac{R\, r_g}{\wp^2}\right){dt}^2
-\frac{\wp^2}{R^2-r_g R+a^2}\,dR^2 -\wp^2d\theta^2-
\nonumber\\\left(a^2+R^2+\frac{R\, r_g
a^2}{\wp^2}\sin^2\theta\right)\sin^2\theta \,d\phi^2+\frac{2 r_g
R \,a}{\wp^2}\sin^2\theta \,d\phi dt,
\label{boyer}
\eea
where $\wp^2=R^2+a^2\cos^2\theta$.

To proceed with the unimodular metric, let?s firstly to do the transformation to a new radial coordinate $r$ considering $R$ as a function $R(r,z)$, where, $z=r\cos \theta$. Writing $dR=\ptl_r R dr +\ptl_z R( dr\cos\theta-r\sin \theta\, d \theta)$, one comes to
\bea
\fl ds^2=\left(1-\frac{R\, r_g}{\wp^2}\right){dt}^2-\wp^2d\theta^2 -
\left(a^2+R^2+\frac{R\, r_g
a^2}{\wp^2}\sin^2\theta\right)\sin^2\theta
\,d\phi^2\nonumber\\+\frac{2 r_g R \,a}{\wp^2}\sin^2\theta
\,d\phi dt-\frac{\wp ^2 ({dr}{\ptl_r R}+\ptl_z R (dr\cos \theta
-d\theta\, r \sin\theta ))^2}{a^2+R^2-R\,r_g}.
\eea
Then let us proceed with the isotopic coordinates $\bm x=\{x,y,z\}$
\begin{eqnarray*}
x=\sqrt{a^2+r^2}\sin\theta\cos\phi,\\
y=\sqrt{a^2+r^2}\sin\theta\sin\phi,
\\
z=r\cos\theta,
\end{eqnarray*}
by writing
\bea
\theta=\arccos \frac{z}{|\bm x|}, \\
\phi=\arctan \frac{y}{x}, \\
r=\frac{1}{\sqrt 2}\sqrt{\bm x^2-a^2+\sqrt{4a^2z^2+(\bm
x^2-a^2)^2}}.\label{rr}
\eea
This leads to the following expression
\bea
\fl ds^2=g_{tt} dt^2+g_1\, (d\bm x\bm J)^2  +g_2\, (d\bm x\bm
x)^2+g_3 (d\bm x\bm x)(d\bm x\bm J)(\bm J\bm x)\\\nonumber+g_4
(\bm J\times \bm x\cdot d\bm x)^2 +g_5 \,dt(d\bm x\cdot\bm x\times
\bm J),
\eea
where $\bm J=\{0,0,1\}$ is an unit vector in the direction of the BH
spin,  $g_i$  are functions of $|\bm x|$, $z^2$,  since $r$ is expressed through $|\bm x|$ and $z$ through (\ref{rr}). As a result, one has:
\bea
g_{tt}=1-\frac{r^2 r_g R}{r^2R^2+a^2
z^2},\nonumber\\
 \fl g_1=-\frac{\left(r^2 R^2+a^2
z^2\right) \left((r^2-z^2) \left(\left(a^2 z^2+r^4\right) \ptl_z
R+a^2 r z \ptl_r R\right)^2-r^8 r_g R+r^8
   R^2+a^2 r^8\right)}{(r^2-z^2)  \left(a^2 r z^2+r^5\right)^2 \left(-r_g R+R^2+a^2\right)},\nonumber\\
\fl g_2=-\frac{r^2 \left(r^2 R^2+a^2 z^2\right) \left(z^2
\left(-r_g R+R^2+a^2\right)+r^2 (r^2-z^2) (\ptl_r
R)^2\right)}{(r^2-z^2) \left(a^2 z^2+r^4\right)^2 \left(-r_g
R+R^2+a^2\right)}, \nonumber\\
\fl g_3=\nonumber\\
\fl \frac{2 r \left(r^2 R^2+a^2 z^2\right) \left((r^2-z^2) \ptl_r
R \left(\left(a^2 z^2+r^4\right) \ptl_z R+a^2 r z \ptl_r R
\right)+r^3( r_g z
   R- z R^2-a^2 z)\right)}{z (r^2-z^2) \left(a^2 z^2+r^4\right)^2 \left(r_g
   R-R^2-a^2\right)},\nonumber\\
   \fl g_4=-\frac{r^2 \left(a^2 \left(r^2+z^2\right) R^2+a^2 r_g (r^2-z^2) R+r^2 R^4+a^4
   z^2\right)}{\left(a^2+r^2\right)^2 (r^2-z^2) \left(r^2 R^2+a^2 z^2\right)},\nonumber\\
   g_5=-\frac{2 a r^2 r_g R}{\left(a^2+r^2\right) \left(r^2 R^2+a^2
   z^2\right)}.\nonumber
\eea
Vectors $\bm N$ and $\bm P$ appearing in (\ref{interv1}) are $\bm
P=0$ and $\bm \gamma\bm N=-g_5 \,\bm x \times \bm J/2$. The FVT theory implies the
restriction $\nabla(\nabla\cdot \bm N)=0$ to vector $N$ because it
is a Lagrange multiplier when the class of the metrics is restricted
by (\ref{interv1}). This restriction is evidently satisfied here
by virtue of $\nabla\cdot \bm N=0$.

The corresponding 3-metric tensor $\gamma_{\mu\nu}$ is
\be
\fl \bm \gamma= g_1\, \bm J\otimes\bm J  +g_2\, \bm x\otimes\bm
x+\frac{g_3}{2} (\bm x\otimes\bm J+\bm J\otimes \bm x)(\bm J\bm
x)+g_4 (\bm J\times \bm x)\otimes (\bm J\times \bm x),
\label{metr}
\ee

 Transformation properties of the
basic vectors are given in a Table \ref{tt}. The  metric tensor
considered (\ref{metr}) is parity and time reversal conserved
quantity. In principle, P-, T- and PT- symmetries are violated in a
nature, but here we restrict ourself only parity and time-reversal invariant
case considering the transformation of only radial coordinate $R(r,z)$.
\begin{table}[t]
\caption{\label{tt}Parity and time-reversal symmetry of the basic
vectors.}
\begin{indented}
\item[]\begin{tabular}{@{}llll} \br
&$\bm x$& $\bm J$&$\bm J\times\bm x$\\
\mr
P-&-        &+&$~~$-\\
T-&+        &-&$~~$-\\
PT-&-&-&$~~$+\\
\br
\end{tabular}
\end{indented}
\end{table}

\begin{figure}[htbp]
\centerline{\includegraphics[width=8cm]{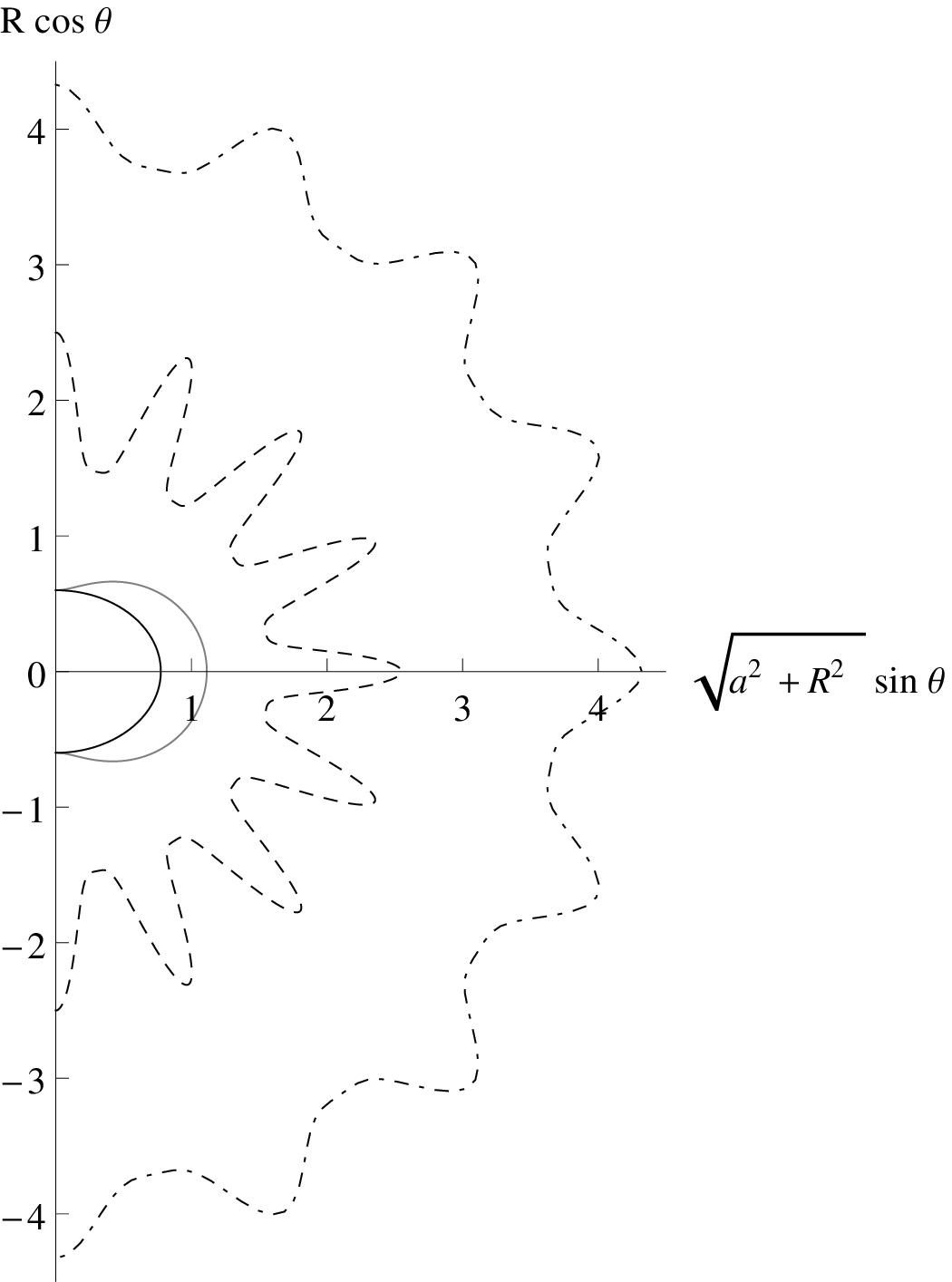}} \caption {The
form of the transformation coordinate surface $R(r,r\cos \theta)$
at $r=0$ - dashed line, at $r=5$ - dashed-dotted line. Horizon and
ergosphere are shown by solid black and gray curves respectively.
Units $r_g=1$ is used.}
\label{surf}
\end{figure}

Conformally unimodular metric (\ref{interv1}) requires that the
coefficient under $dt^2$ in the third degree has to equal the determinant of
the spatial metric (\ref{metr}). This gives the equation for the
function $R(r,z)$ in the form of
\be
\fl\ptl_r R+\frac{z}{r}\ptl_z R=\frac{\left(a^2
z^2+r^4\right)\left(a^2+R^2-r_g\,R \right)^{1/2}  \left(r^2
(R-r_g) R+a^2 z^2\right)^{3/2}}{\left(a^2 z^2+R^2 r^2\right)^2
\left(a^4 z^2+a^2 R^2 \left(r^2+z^2\right)+a^2 r_g R
\left(r^2-z^2\right)+R^4 r^2\right)^{1/2}}.~
\label{eqq}
\ee
From this point, the time variable becomes a conformal time $t=\eta$ of
Eq. (\ref{interv1}). To obtain solution of Eq. (\ref{eqq}), first, it is  convenient to consider the
equation containing an arbitrary function $G$

\be
\ptl_r R+\frac{z}{r}\ptl_z R=G(r,z),
\ee

\noindent which has the formal solution
\be
R(r,z)=\int_0^r
G\left(\xi,\xi\frac{z}{r}\right)d\xi+S\left(\frac{z}{r}\right),
\ee
where $S(\cos\theta)\equiv \mathcal
R(\theta)=R(r,r\cos\theta)\biggr|_{\,r\rightarrow 0}$ is an arbitrary
function determining the boundary condition at $r=0$. Thus, the final
integral equation takes the form
\bea
\fl R(r,r\cos\theta)=\mathcal R(\theta)+\nonumber\\
\fl~~~~~\int_0^r\frac{\left(a^2
\cos^2\theta+\xi^2\right)\left(a^2+R^2-r_g\,R \right)^{1/2}
\left(R(R-r_g)+a^2 \cos^2\theta\right)^{3/2}}{\left(a^2
\cos^2\theta+R^2\right)^2 \left(a^4 \cos^2\theta+a^2 R^2
\left(1+\cos^2\theta\right)+a^2 r_g R
\sin^2\theta+R^4\right)^{1/2}}d\xi,
\label{fineq}
\eea
where it is implied that the function $R$ in the integrating
expression has arguments $R(\xi,\xi\cos\theta)$. Eq.
(\ref{fineq}) allows a numerical solution obtained
iteratively in the form of the Neumann series. The zero-order
approximation is $R=\mathcal R$ and implies that $R$ does not depend on $r$.
Values in the fractional degrees in the integrating expression
should be positive. This restricts $\mathcal R(\theta)$,
namely, $a^2+\mathcal R^2-r_g\,\mathcal R>0$, and $\mathcal
R(\mathcal R-r_g)+a^2 \cos^2\theta>0$, where last inequality corresponds to the ergosphere.

It means returning to a procedure discussed for the Schwarzschild solution when one chooses an arbitrary surface $\mathcal
R(\theta)$ that surrounds the ergosphere and then shirks this surface into a point by the coordinate transformation $R(r,r\cos\theta)$ which satisfies the integral equation (\ref{fineq}). In   Fig.
\ref{surf}, the results of numerical solution are shown, where $
\mathcal R(\theta)=r_g\left(3/2+\cos^4(8\theta)\right)$ and $a=0.49\, r_g$
are taken as an example.

A point-like object is the idealization of the real compact astrophysical object. The real object is made of real matter with the interaction properties, including parity and time-reversal symmetry. Although such a real entity can be formally described as a point, its physical properties preserve and contribute to the characteristics of such a point-like object. In this sense, the parity and time-reversal symmetry conservation/nonconservation are kinds of eicheon ``hairs'' \cite{rgbh7}. Eicheon acquires ``hairs'' because horizon disappears, which violates conditions of the Robinson theorem.  

For $a>r_g/2$, the singularity is naked even in the gauge (\ref{boyer}), which could give rise to the concept of a ``BH electron'' \cite{car,bur}. It seems that the conformally unimodular gauge is even more suitable for such interpretations, because, besides the naked singularity, there exists no ergosphere, and, possibly, no closed time-like geodesics. To obtain an ``elementary eicheon'', one could choose the surface $\mathcal R(\theta)$ exactly along ergosphere.

\section*{References}

\bibliography{hole}

\end{document}